\documentclass[12pt]{iopart}
\usepackage{graphicx}
\usepackage{cite}
\begin{document}                                                   %
\title[Formation of Image-Potential States at the Graphene/Metal Interface]{%
    Formation of Image-Potential States at the
    Graphene/Metal Interface
}
\author{N.~Armbrust$^1$, J.~G{\"u}dde$^1$, and U.~H{\"o}fer$^{1,2}$}
\address{$^1$ Fachbereich Physik und Zentrum f{\"u}r
Materialwissenschaften, Philipps-Universit{\"a}t, 35032 Marburg,
Germany}
\address{$^2$ Donostia International Physics Center (DIPC),
20018 San Sebasti{\'a}n, Spain}
\ead{Jens.Guedde@physik.uni-marburg.de}

\begin{abstract}                                                   %
 The formation of image-potential states at the interface between a
graphene layer and a metal surface is studied by means of model
calculations.
 An analytical one-dimensional model-potential for the combined
system is constructed and used to calculate energies and wave
functions of the image-potential states at the $\bar{\Gamma}$-point
as a function of the graphene-metal distance.
 It is demonstrated how the double series of image-potential states
of free-standing graphene evolves into interfacial states that
interact with both surfaces at intermediate distances, and finally
into a single series of states resembling those of a clean metal
surface covered by a monoatomic spacer layer.
 The model quantitatively reproduces experimental data available for
graphene/Ir(111) and graphene/Ru(0001), systems which strongly
differ in interaction strength and therefore adsorption distance.
 Moreover, it provides a clear physical explanation for the different
binding energies and lifetimes of the first ($n=1$) image-potential
state in the valley and hill areas of the strongly corrugated
moir\'e superlattice of graphene/Ru(0001).
\end{abstract}
%
%

\maketitle

%
\section{Introduction}                                             %
 The ability to fabricate freestanding graphene, a single atomic
layer of graphite, has raised interesting questions with regard to
its surface electronic structure. Due to the fact that this system
exhibits a mirror plane with two surfaces, the eigenstates of this
system must be either a symmetric or antisymmetric superposition of
the electronic states of each surface.
 This applies also to image-potential states, a class of intrinsic
surface states that exist at all solid surfaces due to the
interaction of an electron in the vacuum with the polarizable
surface.
 This interaction can be described by the classical image-potential
which gives rise to a hydrogen-like Rydberg series of electronic
surface states and resonances \cite{Echeni78jp, Fauster95,
Hofer97sci, Hofer15ss}.
 Consequently, it has been predicted that freestanding graphene
should possess a double Rydberg-like series of image-potential
states of even and odd symmetry \cite{Silkin09prb}. However, there
are no experiments so far on freestanding graphene that provide
clear evidence for these two series.

 Image-potential states have been observed when graphene forms an
interface on metal or semiconductor surfaces where it can be grown
with remarkable high quality and long-range order \cite{Wintte09ss,
Emtsev09nat}.
 Experiments on graphene/SiC(0001) using scanning-tunneling
spectroscopy (STS) \cite{Bose10njp, Sandin10apl} and two-photon
photoemission spectroscopy (2PPE) \cite{Takaha14prb} have revealed a
splitting at least of the first $(n=1)$ image-potential state.
 It has been argued that these states are remnants of the first even
and odd state of freestanding graphene even if its mirror symmetry
is in fact broken due to the presence of the substrate.
 In contrast to these findings, only a single series of
image-potential states has been observed on weakly interacting
graphene/metal systems \cite{Niesner12prb, Nobis13prb, Niesner14jp,
Craes13prl}.
 For other metals, like Ni, Pd, Rh, and Ru, the interaction between
graphene and the metal can be much stronger. Together with the
lattice mismatch, this typically leads to a pronounced buckling of
the graphene layer forming a moir\'e superstructure with large
periodicity \cite{Preobr08prb, Wintte09ss}.
 Graphene/Ni(111) represents an exception with an almost vanishing
lattice mismatch. Despite the strong interaction, this results in
the growth of a flat graphene layer at a distance of $d_{\rm
g}=2.1$~{\AA} \cite{Parrei14prb}.
 For g/Ru(0001), on the other hand, the corrugation is particularly
pronounced and the hexagonal moir\'e superlattice is divided into
strongly interacting valley (\emph{L}) areas with a distance of
$d_{\rm g}=2.2$~{\AA} and weakly interacting hill (\emph{H}) areas
with $d_{\rm g}=3.7$~{\AA} \cite{Wang08pccp, Moritz10prl}.
 The observation of field-emission resonances with different energies
in the \emph{L} and \emph{H} areas by STS raised a controversial
debate about the assignment of the lowest members of the series of
Stark-shifted image-potential states \cite{Borca10prl,
Zhang10prlComm, Borca10prlReply, Zhang10jp}.
 Time and angle-resolved 2PPE experiments \cite{Armbrust12prl}
support the general assignment of Borca \emph{et al.}
\cite{Borca10prl}. Based on the 2PPE results, it has been suggested
that the series of image-potential states is slightly decoupled from
the Ru substrate in the \emph{L} areas, whereas the first $(n=1')$
image-potential state in the \emph{H} areas has a substantial
amplitude below the graphene hills. This explains the larger binding
energy, shorter lifetime, and higher effective mass of this
interfacial state which is closely related to the interlayer state
of graphite and of other layered materials \cite{Silkin09prb,
Blase95prb, Cuong14jpcm}.

 Different model potentials have been developed for the description
of image-potential states of freestanding graphene and
graphene/metal systems. Full {\it ab initio} approaches are not
suited for this purpose because the accurate description of the
long-range image force is too costly.
 Silkin {\it et al.} constructed a hybrid potential for freestanding
graphene, which combines a potential, derived from a self-consistent
local density (LDA) calculation for the description of the
short-range properties and an image-potential tail for the proper
description of the asymptotic long-range properties
\cite{Silkin09prb}.
 De~Andres \emph{et\,al.} used an alternative approach that is based
on the dielectric response of the graphene and requires only a
minimal set of free parameters \cite{Andres14njp}. For the
description of graphene/metal systems, however, either the metal
\cite{Andres14njp}, the graphene layer \cite{Paglia15prb} or both
\cite{Zhang10jp} has been represented only by simple effective
barriers characterized by empirical parameters that are fitted to
the experimental results on the binding energies of the specific
system.

 Here, we present an analytical one-dimensional model-potential that
makes it possible to calculate wave functions and binding energies
of image-potential states at the $\bar{\Gamma}$-point for different
graphene/metal systems that are characterized by specific
graphene-metal distances.
 For this purpose we combine potentials for a realistic description
of the projected metal band gap, the image-potential of the metal,
and the potential of the graphene for an arbitrary graphene-metal
distance.
 Furthermore, we include the doping of the graphene layer by
considering the work function difference between graphene and the
metal as well as corrections due to higher-order image-charges.
 All parameters of the model potential are determined from the
properties of the bare metal surfaces and the freestanding graphene
and no free parameters are used for the calculations at different
graphene-metal distances.
 This distance has a strong impact on the properties of the
image-potential states.
 We show how the double Rydberg-like series of freestanding graphene
evolves to a single series when a flat graphene layer approaches the
metal surface as realized for the weakly interacting g/Ir(111) and
the strongly interacting g/Ni(111), which exhibit distinct different
graphene-metal separations.
 The observation of two series can be explained for corrugated
graphene layers with different local graphene-metal separations as
demonstrated for g/Ru(0001).
 The wave functions of the two series of this system are found to
show significant differences in their symmetry, binding energy and
probability density close to the Ru(0001) surface.
 The latter determines the electronic coupling to the metal bulk and
can be connected to the experimentally observed difference of the
inelastic lifetimes of the corresponding states with the same
quantum number in the \emph{L} and \emph{H} areas.

\section{Analytical Model-Potential}                               %
 For the description of the graphene/metal interface we have
constructed a one-dimensional model-potential
\begin{equation}
    V(z) = V_{\rm m}(z) + V_{\rm g}(z-d_{\rm g}) + V_{\rm \Phi}(z) +\delta V(z),
    \label{eq_gpot_v}
\end{equation}
which is composed of a metal potential $V_{\rm m}(z)$ and a graphene
potential $V_{\rm g}(z)$, where $d_{\rm g}$ is the spatial
separation of the graphene layer in the $z$-direction with respect
to the position of the last metal surface atom at $z=0$.
 $V_{\rm \Phi}(z)$ and $\delta V(z)$ are corrections which consider
the work function difference between graphene and the metal and the
influence of higher-order image-charges, respectively. We note that
our potential does not contain any free parameters for the fitting
to experimental binding energies of the combined system.

 For the modeling of the metal, we use the well-established
one-dimensional analytical potential introduced by Chulkov \emph{et
al.} \cite{Chulkov99ss2}.
 This potential describes the bulk by the two-band model of nearly
free electrons and matches the asymptotic image-potential in order
to achieve a total potential $V_{\rm m}(z)$ that is continuously
differentiable for all $z$.
 Its free parameters [see~\ref{sec_m_pot}] are fitted to experimental
data on the size and position of the projected band gap as well as
on the binding energies of the image-potential states of the clean
metal.
 The limitations of the two-band model for the description of the
d-bands in metals like Ru, Ir, and Ni are not important in the
framework of the present work because the properties of the
image-potential states are most sensitively dependent on the
surface-projected bulk band structure in the vicinity of the vacuum
level $E_{\rm vac}$.

 The graphene layer is modeled by a parameterized analytical
potential $V_{\rm g}(z)$ which has been fitted to the numerically
calculated "LDA+image tail" hybrid potential developed for the
description of freestanding graphene by Silkin \emph{et al.}
\cite{Silkin09prb}.
 The binding energies of the image-potential states for freestanding
graphene, which we obtain by using our analytical potential, agree
well with those obtained with the use of the "LDA+image tail" hybrid
potential [see~\ref{sec_g_pot}].
 Although it is known that the $\sigma$ and $\pi$ bands of graphene
are considerably shifted for strongly interacting graphene/metal
systems \cite{Brugger09prb, Varykhalov08prl}, we neglect this change
of the electronic structure by using the same potential for all
graphene-metal distances since this change has only a minor
influence on the image-potential states at the $\bar{\Gamma}$-point,
where the $\sigma$ and $\pi$ bands are far from the vacuum energy
\cite{Silkin09prb}.

\begin{figure}[t]
    \begin{indented}
        \item[]
        \includegraphics[width = 0.6\columnwidth]{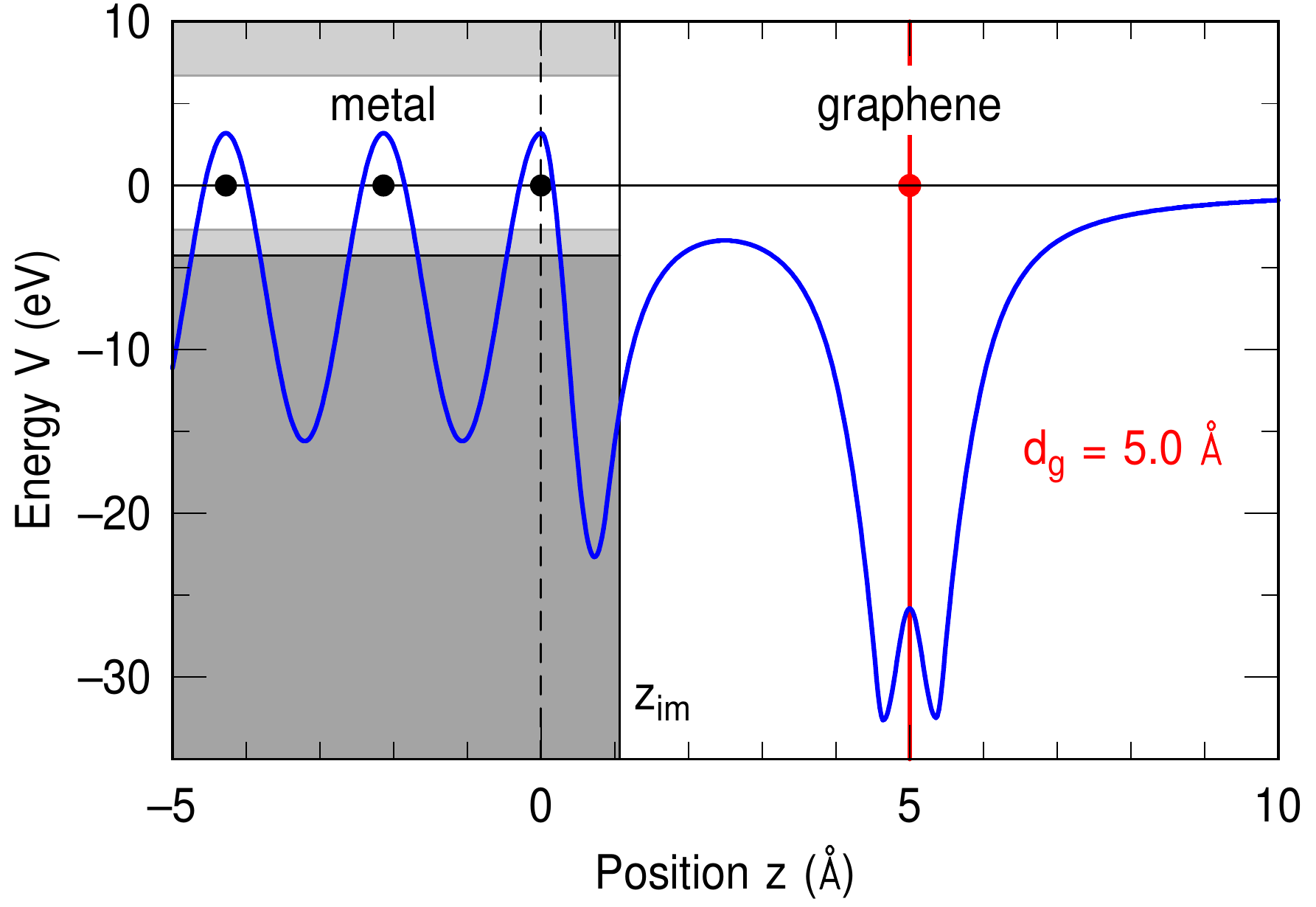}
        \caption[]{One-dimensional model-potential $V(z)$
            for graphene/Ru(0001) at a graphene-metal distance of
            $d_{\rm g} = 5.0$~{\AA}.
            Black and red circles denote the position of Ru and
            C atomic layers, respectively.
            \label{fig_gpot}}
    \end{indented}
\end{figure}

 $\delta V(z)$ describes the effect of higher-order image charges
which arise at $0<z<d_{\rm g}$ due to infinite multiple reflections
of the primary image charges at the metal surface and the graphene
layer.
 The latter can be regarded as metallic because of the efficient
screening of lateral electric fields by the \emph{sp}$^2$ layer
\cite{Brugger09prb}.
 The classical image-potential between two metal surfaces with
distance $d_{\rm g}$ is given by
\begin{equation}
     V_{\rm im}(z) = -\frac{1}{4z} - \frac{1}{4(d_{\rm g}\!-\!z)}
                    +\frac{1}{4d_{\rm g}} + \frac{1}{4d_{\rm g}}-\frac{1}{4(d_{\rm g}\!+\!z)}- \frac{1}{4(2d_{\rm g}\!-\!z)}+\ldots\label{eq_multi}.
\end{equation}
 The first two terms in \eref{eq_multi} represent the image-potential
in atomic units caused by of the primary image-charges from each
metallic surface.
 The following terms result from subsequent multiple reflections.
 By reordering, the latter can by written as
\begin{equation}
     \delta V(z) = \frac{1}{4}\sum_{k=1}^{\infty}\frac{2}{kd_{\rm g}}
                  -\frac{1}{kd_{\rm g}+z}
                  -\frac{1}{(k+1)d_{\rm g}-z}\ .
     \label{eq_delta_v}
\end{equation}
 This series converges to a net repulsive contribution. It
increases the potential maximum in the center between the graphene
layer and the metal surface (\fref{fig_gpot}) from $V_{\rm
im}(z\!=\!d_{\rm g}/2)=-1/d_{\rm g}$ to $V_{\rm im}(z\!=\!d_{\rm
g}/2)=-\ln(2)/d_{\rm g}$, which is the classic value of the
self-energy of a charged particle located in the gap between two
metals \cite{Sols87prb}.
 For decreasing distance $d_{\rm g}$, the potential maximum is
lowered.
 However, even at $d_{\rm g}=2.2$~{\AA}, as found in \emph{L} areas
of g/Ru(0001), $V_{\rm im}(d_{\rm g}/2)=-4.54$~eV remains above the
Fermi level.
 A simplified description of the potential between metal and graphene
by an effective quantum well with a large depth of 4.8~eV {\em
below} the Fermi level, as proposed by Zhang \emph{et al.}
\cite{Zhang10jp}, thus results in unrealistically large binding
energies of the image-potential states. This is the main origin of
the discrepancy in the identification of the first field-emission
resonance observed in the STS spectra \cite{Borca10prl,
Zhang10prlComm, Borca10prlReply, Zhang10jp}.

 The correction $V_{\rm \Phi}(z)$ considers the charge transfer
between metal and graphene. It is approximated by a linear
interpolation of the work function difference $\Delta\Phi$ between
the clean metal surface and the combined system:
\begin{equation}
     V_{\rm \Phi}(z)=\Delta\Phi\left(1-\frac{z}{d_{\rm g}}\right)\quad 0 < z \leq d_{\rm g}.
\end{equation}

 On the basis of the model potential $V(z)$, which is depicted for
the example g/Ru(0001) at a graphene separation $d_{\rm g} =
5.0$~{\AA} in \fref{fig_gpot}, the wave function of the
image-potential states and the corresponding binding energies
$E_{\rm n}$ with respect to the vacuum level $E_{\rm vac}$ have been
obtained by solving the one-dimensional Schr{\"o}dinger equation
numerically by Numerov's method.
 By extending the model potential to more than $100$~{\AA} into the
metal bulk and to more than $1000$~{\AA} into the vacuum, we made
sure that the results do not depend on the system's extension.
 Even if we halve these intervals, the binding energies of the
image-potential states with quantum numbers $n\le 4$ change by much
less than $0.1$~meV.
%
\begin{figure}[t]
    \begin{flushright}
        \includegraphics[width = \columnwidth]{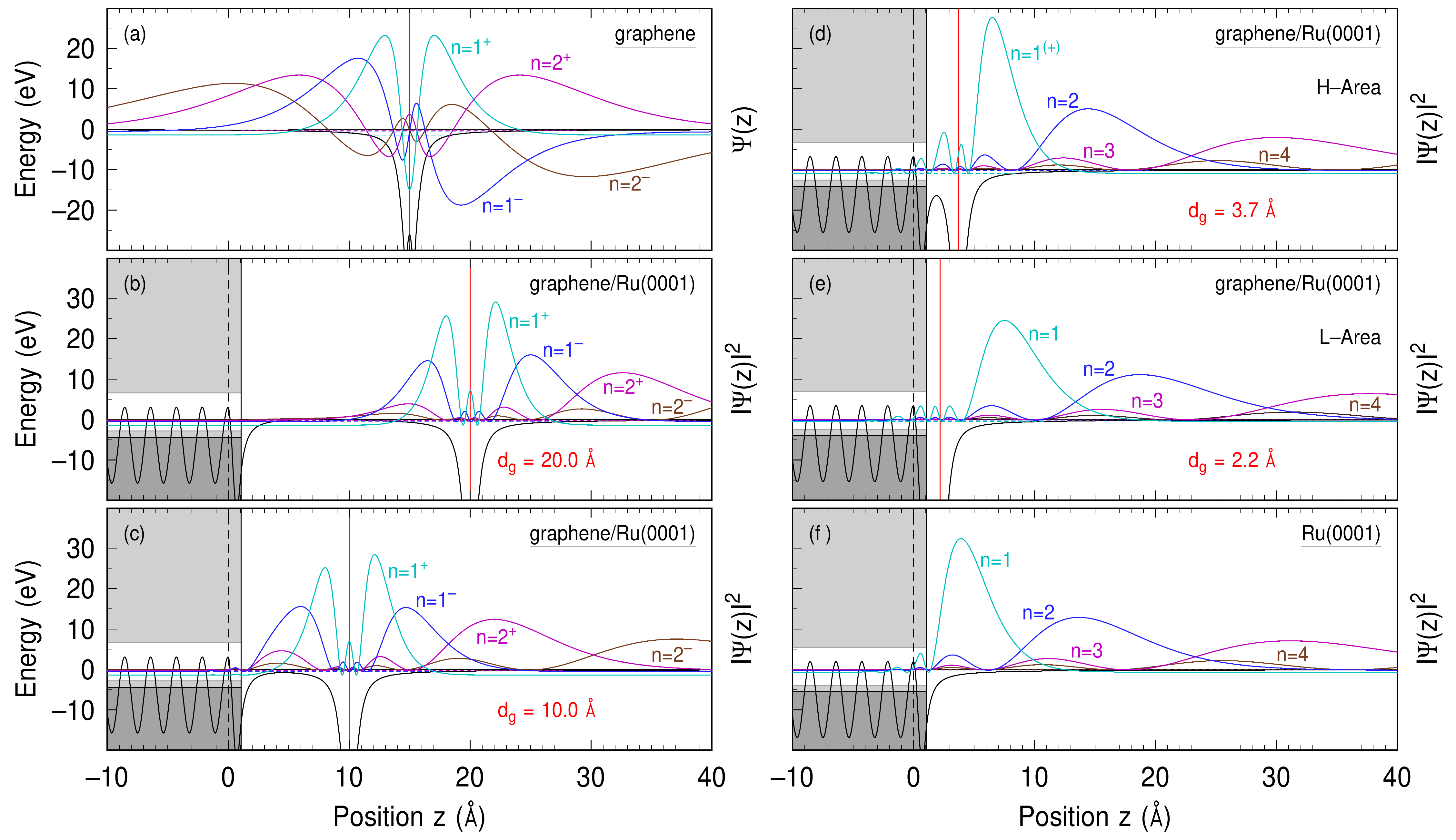}
        \caption[]{(a) Wave functions $\Psi(z)$ of the first two
            members of the symmetric ($+$) and antisymmetric ($-$)
            image-potential states of freestanding graphene,
            respectively.
             (b)-(f) Probability densities
            $|\Psi(z)|^{2}$ of the image-potential states at the
            graphene/Ru(0001) interface for graphene-Ru separations
            $d_{\rm g}$ of $20$~{\AA} (b),
            $10$~{\AA} (c),
            $3.7$~{\AA} (d), \emph{H} areas) and
            $2.2$~{\AA} (e), \emph{L} areas) and at the
             clean Ru(0001) surface (f).
             The positions of the uppermost Ru atomic layer and the
            graphene layer are shown by vertical black dashed and
            red solid lines, respectively.
             The solid black line depicts the one-dimensional
            model-potential.
             The projected Ru bulk band structure (gray shaded
            areas) has been extended up to the image-plane
            of the metal
            ($z=z_{\rm im}$).
            \label{fig_probability_densities}
            }
\end{flushright}
\end{figure}
%

\section{Results and Discussion}                                   %

 \Tref{tab_results} summarizes the calculated and experimental
binding energies $E_{n}$ with respect to $E_{\rm vac}$ of the first
two image-potential states for graphene on Ru(0001), Ir(111), and
Ni(111).
 It shows that our model achieves a very good quantitative agreement
with the experimentally observed binding energies which have been
obtained by 2PPE for g/R(0001) \cite{Armbrust12prl} and g/Ir(111)
\cite{Niesner12prb}.
 In particular, the distinct binding energies of the ($n=1$) state
in the \emph{L} and \emph{H} areas of g/Ru(0001) can be reproduced.
 Together with the predicted values for g/Ni(111), this data suggests
that the binding energies of the image-potential states are strongly
correlated with the graphene-metal separation $d_{\rm g}$.
 In order to emphasize this correlation, we will discuss in this
section the systematics of the formation of image-potential states
at different graphene-metal separations for the example of the
strongly corrugated g/Ru(0001) because its \emph{L} and \emph{H}
areas exhibit distinct graphene-metal separations, that are typical
for strongly and weakly interacting graphene/metal systems,
respectively \cite{Niesner14jp}.

\Table{Calculated binding energies $E_{n}$~(eV) with respect to the
        vacuum level of the first two image-potential states ($n=1$)
        and ($n=2$) in the \emph{L} and \emph{H} areas of
        graphene/Ru(0001) and on almost flat graphene/Ir(111) and
        graphene/Ni(111).
         The results of the one-dimensional model-potential are
        compared to the experimental values obtained by
        2PPE \cite{Armbrust12prl, Niesner12prb}.
         $d_{\rm g}$~({\AA}) denotes the respective graphene-metal
        separations.
        \label{tab_results}}
\br
 &&\centre{2}{1D-Potential} & \centre{2}{Experiment}\\
 \ns
&&\crule{2}&\crule{2}\\
& \centre{1}{$d_{\rm g}$} & \centre{1}{$E_{1}$} & \centre{1}{$E_{2}$} & \centre{1}{$E_{1}$} & \centre{1}{$E_{2}$} \\
\mr
g/Ru(0001) H \cite{Armbrust12prl} & 3.7 & 0.95 & 0.24  & $0.80\pm0.04$  & $0.18\pm0.06$\\
g/Ru(0001) L \cite{Armbrust12prl} & 2.2 & 0.47 & 0.16  & $0.41\pm0.04$  & $0.18\pm0.06$\\
g/Ir(111) \cite{Niesner12prb}     & 3.4 & 0.88 & 0.23  & $0.83\pm0.02$  & $0.19\pm0.02$ \\
g/Ni(111)                         & 2.1 & 0.41 & 0.15 \\
\br
\endTable

 The \emph{L} and \emph{H} areas of g/Ru(0001) form a hexagonal
moir\'{e}-superlattice with a lateral edge length of approximately
$30$~{\AA} of its $12.5\!\times\!12.5$-subunits \cite{Martoc08prl,
Martoc10njp}.
 Within our one-dimensional model we can treat both areas separately
because the lateral confinement due to the finite size of each area
will reduce the binding energy of the image-potential states only
slightly, as can be estimated from the area size of a half subunit
$A\approx 420$~{\AA}$^2$ to $\hbar^{2}\pi^{2}/2m_{\rm eff}A\approx
43$~meV, where $m_{\rm eff}=2.1\ m_{\rm e}$ with respect to the free
electron mass $m_{\rm e}$ \cite{Armbrust12prl}.
 Since charge-transfer between the metal and graphene leads to a
doping of graphene \cite{Brugger09prb, Sutter09nl}, we model the
work function of g/Ru(0001) as an empirical function of the
graphene-Ru separation $d_{\rm g}$ by $\Phi_{\rm g/Ru}(d_{\rm g}) =
\Phi_{\rm g}-\xi/d_{\rm g}$, which makes it possible to vary $d_{\rm
g}$ systematically for the model calculations.

 Setting $\xi=1$~eV\,{\AA} matches the work function $\Phi_{\rm
H}=4.24$~eV in the \emph{H} areas \cite{Armbrust12prl} as well as
the $0.24$~eV lower value in the \emph{L} areas \cite{Brugger09prb}.
For large separation, it converges to $\Phi_{\rm g}=\Phi(d_{\rm
g}=\infty)= 4.48$~eV \cite{Giovan08prl} of freestanding graphene.

\begin{figure}[t]
    \begin{indented}
        \item[]
        \includegraphics[width = 0.6\columnwidth]{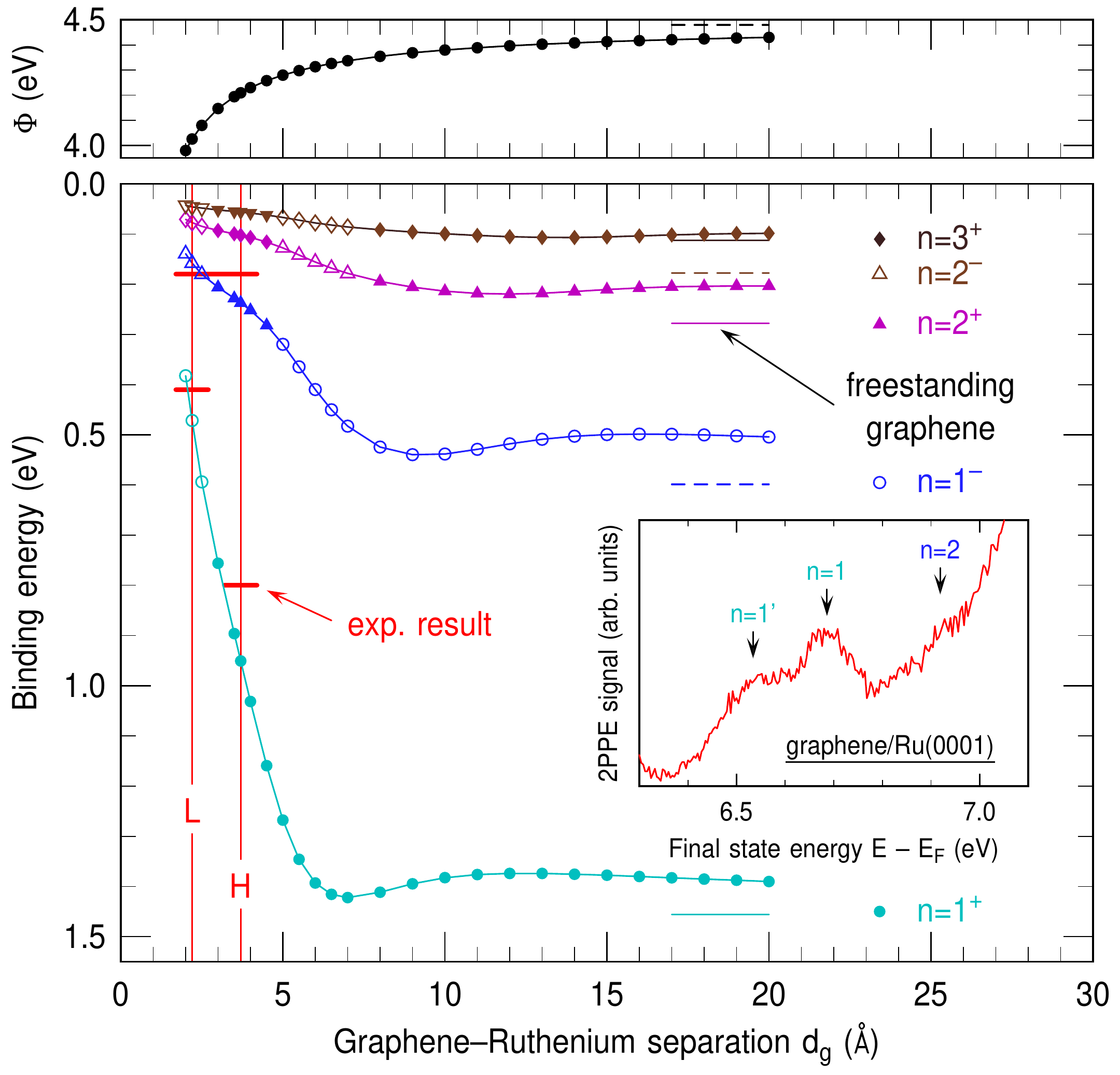}
        \caption[]{Work function and
            calculated binding energies
            with respect to the vacuum level of the first
            four members of image-potential states at graphene/Ru as
            a function of the separation $d_{\rm g}$ between the
            graphene layer and the Ru(0001) surface.
             Closed (open) symbols denote states with a symmetry at
            $z=d_{\rm g}$ which is close to the symmetry of the even
            (odd) image-potential states of free-standing graphene.
             The binding energies of the first two even ($n=1^+$,
            $2^+$) and odd ($n=1^-$, $2^-$) image-potential states
            of freestanding graphene are depicted by solid and
            dashed horizontal lines, respectively.
             The vertical lines indicate the graphene-Ru separations
            of the \emph{L} and \emph{H} areas.
             The inset shows the experimental 2PPE spectrum
            adapted from \cite{Armbrust12prl} which
            averages over a large number of \emph{L} and \emph{H}
            areas.
            \label{fig_energy_2ppe}}
    \end{indented}
\end{figure}

 \Fref{fig_probability_densities} shows the calculated probability
densities of the image-potential states for different separations
$d_{\rm g}$ between the graphene layer and the Ru(0001) surface as
well as for the separate systems.
 The corresponding binding energies are depicted in
\fref{fig_energy_2ppe}. For freestanding graphene, the two series
with even ($+$) and odd ($-$) symmetry can clearly be assigned from
the amplitude of their wave functions at $z=d_{\rm g}$ where the odd
states have a node, whereas the even states have an antinode.
 The presence of the metal substrate breaks the mirror symmetry of
the freestanding graphene and the parity of the image-potential is,
in principle, no longer a good quantum number.
 At a graphene-metal separation as large as $d_{\rm g}=20$~{\AA},
however, the influence of the substrate on the lowest members
($n=1^{+}, 1^{-}$) of the image-potential states is still weak due
to the rather small spatial extent of their wave functions
perpendicular to the surface.
 Therefore, their probability densities almost resemble those of
freestanding graphene.
 We point out, that even for large $d_{\rm g}$, only
image-potential states that originate from graphene and no states
that are bound in the potential well formed by the image-potential
of the metal and its surface barrier are found.
 This is related to the large difference in work function between
graphene and Ru(0001) ($\Phi_{\rm Ru}=5.51$~eV
\cite{Armbrust12prl}), which results in an additional repulsion from
the Ru surface. This contributes to a reduction of the binding
energies with respect to those of freestanding graphene.
 The latter are depicted by dashed lines in \fref{fig_energy_2ppe}.
 The same argument applies for the Ir(111) ($\Phi_{\rm Ir}=5.79$~eV
\cite{Niesner12prb}) and the Ni(111) ($\Phi_{\rm Ni}=5.27$~eV
\cite{Andres15jp}) surface.
 Beside the work function difference, the repulsion due to the
surface barrier of the metal contributes additionally to a reduction
of the binding energies with respect to freestanding graphene.
 At $d_{\rm g}=20$~{\AA}, this applies particularly
to the higher members ($n=2^{\pm}$, $3^{\pm}$, \ldots) because of
their more extended wave functions, which are already considerably
perturbed by the repulsion due to the surface barrier.
 This also results in an asymmetry and phase shift of the wave
functions with respect to those of freestanding graphene.
 The wave function of the state labeled as
$n=2^-$ at $d_{\rm g}=20$~{\AA}, for example, has now almost an
antinode at $z=d_{\rm g}$ and more closely resembles the wave
function of the ($n=3^+$)-state of freestanding graphene as denoted
by the corresponding symbol in \fref{fig_energy_2ppe}.

 At intermediate separations, the properties of the image-potential
states start to be governed by the image-potentials of both
surfaces, and the attractive image-potential of the metal partly
compensates for the repulsion due to the surface barrier and the
work function difference.
 The attraction of the metal leads to a maximum of the binding energy
of each state at specific separations as can be seen in
\fref{fig_energy_2ppe}.
 This is also reflected by the form of the wave functions shown in
\fref{fig_probability_densities}.
 At $d_{\rm g}=10$~{\AA}, for example, the symmetric ($n=1^+$)-state
is still almost unaffected while the more extended wave function of
the ($n=1^-$)-state is slightly attracted towards the metal surface
which results in an increase of the probability density in between
the metal surface and the graphene layer compared to larger
separations.
 For comparison we have also made simplified calculations (not shown)
where we described the metal as a simple barrier without an
image-potential similar to the model used in \cite{Andres14njp}.
 In this case, we do not observe this attraction and all
image-potential states are always repelled when the graphene layer
approaches the metal surface.

 At small graphene-metal separations ($d_{\rm g}< 5$~{\AA}), the
repulsion due to the surface barrier of the metal dominates and also
the phase of the ($n=1^-$) wave function shifts and the latter
starts to resemble that of the higher-lying ($n=2^+$)-state of
freestanding graphene.
 In particular at a separation of $d_{\rm g}=3.7$~{\AA}, which
corresponds to the \emph{H} areas of g/Ru(0001), the vacuum part
($z>d_{\rm g}$) of all wave functions almost resembles the
corresponding part of the symmetric series of freestanding graphene.
 This can be regarded as a transition from the double series of
image-potential states of freestanding graphene to one single
series, omitting the differentiation with regard to the parity.
 Since only the former ($n=1^+$)-state maintains a certain similarity
to the corresponding state on freestanding graphene, we denote this
fact by writing $+$ in parentheses in the \emph{H} areas.
 At a separation of $d_{\rm g}=2.2$~{\AA}, which corresponds to the
\emph{L} areas of g/Ru(0001), all wave functions are further
repelled from the metal surface.
 Even if the phase of the wave functions now correspond to the phase
of the odd series of states on free standing graphene, the mirror
symmetry has completely vanished, which is indicated by omitting the
parity for the whole series.
 Now, all wave functions almost resemble the form of those on the
clean Ru surface with one layer of atoms atop, as can be seen by
comparing \fref{fig_probability_densities}(e) and (f).
 In comparison to the clean Ru surface, however, the
spatial extent of the wave functions in the vacuum is larger, which
indicates a stronger decoupling from the surface.
 In the \emph{H} areas, on the other hand, a considerable amount of
the ($n=1^{(+)}$) wave function still fits in between the graphene
and the Ru surface and is therefore more strongly coupled to Ru as
compared to the \emph{L} areas.
 This state is closely related to the interlayer state of graphite
\cite{Silkin09prb}.

 As depicted by the different symbols in \fref{fig_energy_2ppe}, the
change of the wave function's character upon approaching the
graphene layer to the metal surface is also reflected by the change
of the binding energy with respect to the vacuum level.
 Whenever the binding energy of the image-potential states of the
entire system becomes smaller compared to the next energetic
adjacent state of freestanding graphene, the local symmetry at
$z=d_{\rm g}$ is closer to the latter.
 At $d_{\rm g}=20$~{\AA}, for example, the binding energies of the
first, second, and third state of the entire system are larger
compared to the second, third, and fourth state of freestanding
graphene.
 Consequently, their local symmetry at $z=d_{\rm g}$ is still close
to that of the first ($n=1^+$), second ($n=1^-$), and third
($n=2^+$) state of freestanding graphene.
 The energy of the fourth state of the entire system, however, is
already smaller compared to the fifth state ($n=3^+$) of
freestanding graphene.
 Its local symmetry at $z=d_{\rm g}$ more closely resembles the
symmetry of that state and not of the fourth state ($n=2^-$) of
freestanding graphene.
 The energy of this state is therefore drawn as a closed symbol which
denotes states of even (local) symmetry.
 For decreasing graphene-metal separation, the binding energies of
the image-potential states increase slightly at first due to the
attractive image-potential of the metal.
 At separations found in real graphene/metal systems ($d_{\rm g}<
5$~{\AA}), however, the repulsion due to the surface barrier of the
metal dominates and results in a rapid decrease of the binding
energies upon further approach.
 The sequence of the relative energy spacings is, however, rather
insensitive towards $d_{\rm g}$ and resembles a scaled Rydberg
series with decreasing energy separation for increasing quantum
number for all $d_{\rm g}$.

 The absolute change of the binding energy as a function of $d_{\rm
g}$ is most pronounced for the first image-potential state.
 It decreases from more than 1.4~eV at large $d_{\rm g}$ to 0.95~eV
at the separation found in the \emph{H} areas of g/Ru(0001) and
finally to 0.47~eV at the separation found in the \emph{L} areas.
 The difference between the binding energies with respect to $E_{\rm
vac}$ in the \emph{H} and \emph{L} areas of 0.48~eV is twice as
large as the difference between the local work functions of 0.24~eV
\cite{Brugger09prb}.
 The first image-potential state in the \emph{H} areas has
therefore a smaller energy with respect to the Fermi level $E_{\rm
F}$ when compared to the the corresponding state in the \emph{L}
areas.
 This confirms the assignment of the peaks in the 2PPE spectrum of
\cite{Armbrust12prl} which is reproduced in the inset of
\fref{fig_energy_2ppe}.
This assignment is also in agreement with density functional
theory (DFT) calculations by Borca \emph{et al.} \cite{Borca10prl}
which shows that the results of our model calculations are robust
also for small distances where DFT can explicitly describe
the chemical interaction between graphene and the metal.
For large distances and higher quantum numbers the evolution
of the binding energies with distance depicted in \cite{Borca10prl}
considerably differs from our results which we attribute
to the fact that DFT cannot properly consider the long-range image-force.

 For the second image-potential state, the calculated difference
between binding energies with respect to $E_{\rm vac}$ in the
\emph{H} and \emph{L} areas is in the order of the difference in the
local work function, which explains that these two states could not
be resolved separately in the 2PPE spectra.
 The red horizontal lines in \fref{fig_energy_2ppe} depict the
experimental results from \cite{Armbrust12prl} under consideration
of the local work functions in the \emph{H} and \emph{L} areas, and
help visualize the very good agreement between experiment and model
calculation.
 The small deviation for the binding energy of the lowest
image-potential state energy in the \emph{H} areas is reduced if one
considers the additional reduction of the calculated binding energy
due to the lateral localization as discussed above.
 The localization of this state should be stronger in the \emph{H}
areas as deduced from the effective mass which has been found to be
twice as large as the free electron mass \cite{Armbrust12prl}.

 The results of our model calculation make it also possible
to qualitatively interpret the experimentally observed decay
dynamics of electrons in these states, as has been investigated by
time-resolved 2PPE \cite{Armbrust12prl}.
 As shown in \fref{fig_probability_densities}(e), the probability
densities of all image-potential states in the \emph{L} areas mostly
resemble those of the clean Ru(0001) surface but with some degree of
decoupling, i.e.\ the center of gravity of the probability density
is located further away from the metal into the vacuum.
 At first glance, this similarity to the clean Ru surface seems to be
surprising.
 The graphene-metal separation in the \emph{L} areas is, however,
almost identical to the Ru-layer spacing ($d_{\rm Ru}=2.14$~{\AA}
\cite{Pelzer00jp}).
 The Ru surface covered by one metallic-like graphene overlayer
therefore has similar properties with respect to the image-potential
states as the extension of the Ru crystal by one more layer of Ru
atoms.
 At the clean Ru(0001) surface, the lifetimes of electrons in the
first two image-potential states of $\tau_{1}=11$~fs and
$\tau_{2}=57$~fs \cite{Gahl04dr, Berthold00cp} are comparably short
because the decay by inelastic scattering with bulk electrons is
very efficient due to the high density of states of the Ru 4d bands
just below the Fermi level \cite{Berthold00cp}.
 Thus, the slightly enhanced experimental lifetimes of $\tau_{1^{-}}
=16\pm5$~fs and $\tau_{2^{-}} =85\pm13$~fs \cite{Armbrust12prl} are
a direct indication for a certain degree of decoupling in the
\emph{L} areas.
 In the \emph{H} areas, a considerable part of the probability
density of the ($n=1^{(+)}$)-state fits below the graphene layer
close to the Ru surface.
 This enhances the electronic coupling to the Ru bulk states as is
reflected by the shorter lifetime of $\tau_{1^{(+)}}=11\pm8$~fs
\cite{Armbrust12prl}, which is identical to the value found on clean
Ru(0001).

 By adapting $V_{\rm m}(z)$, our model can be easily applied to the
description of graphene layers on other metal substrates.
 With Ir(111) and Ni(111) we chose two examples that represent
limiting cases of weak and strong interaction with the graphene
layer, respectively.
 It has been shown that g/Ir(111) grows with a rather large and
homogeneous separation of $d_{\rm g}=3.4$~{\AA} \cite{Busse11prl}
which is very similar to the separation found in the \emph{H} areas
of g/Ru(0001).
 Graphene and Ni(111) have almost no lattice mismatch.
 This results in the growth of a flat graphene layer on Ni(111),
despite the strong interaction \cite{Parrei14prb}.
 The resulting separation of about $d_{\rm g}=2.1$~{\AA} is very
similar to that found in the \emph{L} areas of g/Ru(0001).
 Binding energies and lifetimes of the image-potential states on
g/Ir(111) have been measured by 2PPE \cite{Niesner12prb} and can be
directly compared with the results of our model calculations. No
experimental data on g/Ni(111) are available, but 2PPE results on
the binding energies for the clean Ni(111) surface \cite{Andres15jp}
make it possible to properly adjust the parameter of the metal
potential and to predict the binding energies for the graphene
covered surface.
 The ferromagnetic coupling of Ni results in a
spin-split surface-projected band structure which is reflected by an
exchange-splitting of the image-potential states.
 The observed splitting of $14\pm 3$~meV is, however, small compared
to the linewidth and can only be observed with spin-resolved
detection \cite{Andres15jp}.
 We therefore neglect the exchange-splitting here and use a
spin-integrated surface-projected band gap extracted from
\cite{Schupp92prb} for our model calculations.
 Because the size and position of the surface-projected bulk band gap
do not differ much between Ru(0001), Ir(111), and Ni(111), the
calculated binding energies depend most sensitively on the
graphene-metal separation $d_{\rm g}$.
 Consequently, the binding energies for g/Ir(111), as well as for
g/Ni(111), are very similar to the results for g/Ru(0001) at the
corresponding separations.
 As shown in \tref{tab_results}, the results of our model calculation
for the binding energies on g/Ir(111) are in excellent agreement
with the experimental data.
 For g/Ni(111) we predict much smaller binding energies compared
to g/Ir(111).
 This is due to the stronger repulsion of the image-potential states
at the smaller separations.
 The binding energies on g/Ni(111) are on the other hand comparable
to those on the \emph{L} areas of g/Ru(111) in correspondence to the
similar separation $d_{\rm g}$.

 Finally, we would like to comment on the formation of
image-potential states at the interface of graphene/SiC, even if our
model is not directly applicable to this semiconducting substrate.
 Although graphene might possess only a weak interaction to SiC, our
results generally show that the shape of the image-potential states
is very sensitive to the substrate induced symmetry break even for
large separations due to their large spatial extent.
 In another 2PPE experiment on graphene/SiC(0001), Shearer
\emph{et al.} \cite{Shearer14apl} indeed observed a single series of
sharp symmetric peaks of image-potential states in agreement with
data shown in \cite{Niesner14jp}.
 Therefore, it seems rather unlikely that the
experimentally found splitting of the first image-potential state on
graphene/SiC \cite{Bose10njp, Takaha14prb} can be interpreted as a
remnant of the mirror symmetry of freestanding graphene with a small
energy separation between these two states.
 The latter is particularly hard to understand because the approach
of a flat graphene layer towards a substrate with a surface barrier
does not change the relative energy spacing of adjacent states,
i.e.\ the energy spacing between the lowest two states is always
larger than the subsequent spacing because the energy sequence of
the Rydberg series is at all distances dominated by the long-range
image-potential.

\section{Conclusion}                                               %

 On the basis of an analytical, one-dimensional potential, we can
quantitatively describe the formation of image-potential states at
the interface between a graphene layer and a metal surface by means
of model calculations.
 By systematic variation of the graphene-metal separation, we have
shown how the double Rydberg-like series of even and odd
image-potential states of freestanding graphene evolves towards a
single series of the compound system when a flat graphene layer is
located at a distance to the metal as found in real graphene/metal
systems.
 This transition is driven by the repulsion from the metal substrate,
which increasingly reduces the mirror symmetry of freestanding
graphene with decreasing separation.
 Even for separations found in weakly interacting systems, we find no
remnant of the distinction between states of even and odd symmetry.
 At these intermediate separations, the first image-potential state
is partly trapped between the metal and the graphene layer. It
attains properties of an interfacial state.
 At distances found in strongly interacting systems, on the other
hand, the wave functions almost resemble those of image-potential
states on a clean metal surface but with a higher degree of
decoupling.
 The repulsion of the image-potential states from the metal surface
is also responsible for a strong reduction of their binding energies
with decreasing graphene-metal separation.
 This explains the distinct differences of the experimentally
observed binding energies of the first ($n=1$) image-potential
states in the \emph{H} and \emph{L} areas of strongly corrugated
g/Ru(0001).
 According to the respective similar separations on weakly
interacting g/Ir(111) and strongly interacting g/Ni(111), we find
respective comparable binding energies which are in excellent
agreement with available experimental data for g/Ir(111).

\ack                                                               %
 We thank Richard H{\"o}fer for valuable contributions to the
multiple-reflection correction of the image-potential and gratefully
acknowledge funding by the Deutsche Forschungsgemeinschaft through
SFB 1083 \emph{Structure and Dynamics of Internal Interfaces}
Project, B6.

\appendix                                                          %
\section{Metal Potential}                                          %
\label{sec_m_pot}                                                  %
%
 For the metal part $V_{\rm m}(z)$ of our model potential we used the
well-established one-dimensional potential introduced by Chulkov
\emph{et al.} \cite{Chulkov99ss2}.
 This parameterized potential is piecewise defined along the surface
normal $z$ by:
\begin{equation}
V_{\rm m}(z) = \cases{
    -A_{10}+A_{1}\cos\left(\frac{2\pi }{d_{\rm m}}z\right)      &$\ z \leq 0$\\
    -A_{20} +A_{2}\cos(\beta z)         &$0 < z \leq z_{1}$\\
    -A_{3}\exp[-\alpha(z-z_{1})]                                &$z_{1} <  z \leq z_{\rm im}$\\
    \;\;\,\frac{\exp[-\lambda(z-z_{\rm im})]-1}{4(z-z_{\rm im})}  &$z_{\rm im} <  z$
    }
    \label{eq_mpot}
\end{equation}
 Here $d_{\rm m}$ is the layer spacing and $z=0$ corresponds to the
position of the surface atom.
 By requiring that $V_{\rm m}(z)$ and its first derivative to be
continuous at the matching points $z_1$ and $z_{\rm im}$, only four
of the ten parameters $A_{1}$, $A_{10}$, $A_{2}$, $A_{20}$, $A_{3}$,
$\alpha$, $\beta$, $\lambda$, $z_{1}$, $z_{\rm im}$ are the
independent of each other.
 It is feasible to choose the offset $A_{1}$ and the amplitude
$A_{10}$ in order to reproduce the energetic position and width of
the surface-projected band gap.
 For Ru(0001), Ir(111) and Ni(111) (spin averaged) these have
been extracted from \cite{Holzwa85ssc,Lindroos86prb1},
\cite{Niesner12prb} and \cite{Schupp92prb}, respectively.
 The parameters $A_{2}$ and $\beta$ have been used to reproduce the
experimental binding energies $E_{n}$ of the image-potential states.
 For Ru(0001) we have fitted these parameters to the experimental
binding energies $E_{1}=0.66$~eV and $E_{2}=0.19$~eV reported by
Gahl \emph{et al.} \cite{Gahl04dr}.
 For Ni(111) we have used data reported by Andres \emph{et al.}
\cite{Andres15jp}. Due to the lack of experimental data on the
image-potential states of clean Ir(111), we have estimated the
binding energies by using the Rydberg formula $E_{n}=0.85\;{\rm
eV}/(n+a)^{2}$ where $a$ is the quantum defect.
 For a $sp$-inverted band gap, the latter can be determined from the
position of the vacuum energy $E_{\rm vac}$ within the projected
band gap \cite{Fauster95}.
 This results in $E_{1}=0.64$~eV and $E_{2}=0.18$~eV.

\Table{Parameters of the metal part $V_{\rm m}(z)$ of the
        model-potentials used for Ru(0001), Ir(111), and Ni(111).
         The work function $\Phi$, the potential offset $A_{10}$ and
        amplitude $A_{1}$ are given in eV, the layer spacing
        $d_{\rm m}$ in {\AA} and the parameter $\beta$ in
        {\AA}$^{-1}$.
        \label{tab_m_pot}}
\br
 & $d_{\rm m}$ & $\Phi$ & $A_{10}$ & $A_{1}$ & $A_{2}$ & $\beta$\\
\mr
Ru(0001)& 2.138 \cite{Pelzer00jp} & 5.51 \cite{Armbrust12prl} & \07.436 & 9.400  &  12.70  & 4.3464 \\
Ir(111) & 2.217 \cite{Needs89jp}  & 5.79 \cite{Niesner12prb}  & \09.511 & 6.188  & \06.50  & 4.6068 \\
Ni(111) & 2.035 \cite{Batche54ac} & 5.27 \cite{Andres15jp}    &  11.259 & 6.503  & \04.91  & 2.3070 \\
\br
\endTable

 The parameter sets used for modelling Ru(0001), Ir(111), and Ni(111)
are collected in \tref{tab_m_pot}.

\section{Graphene Potential}                                       %
\label{sec_g_pot}                                                  %
\begin{figure}[tb]
    \begin{indented}
        \item[]
        \includegraphics[width = 0.6\columnwidth]{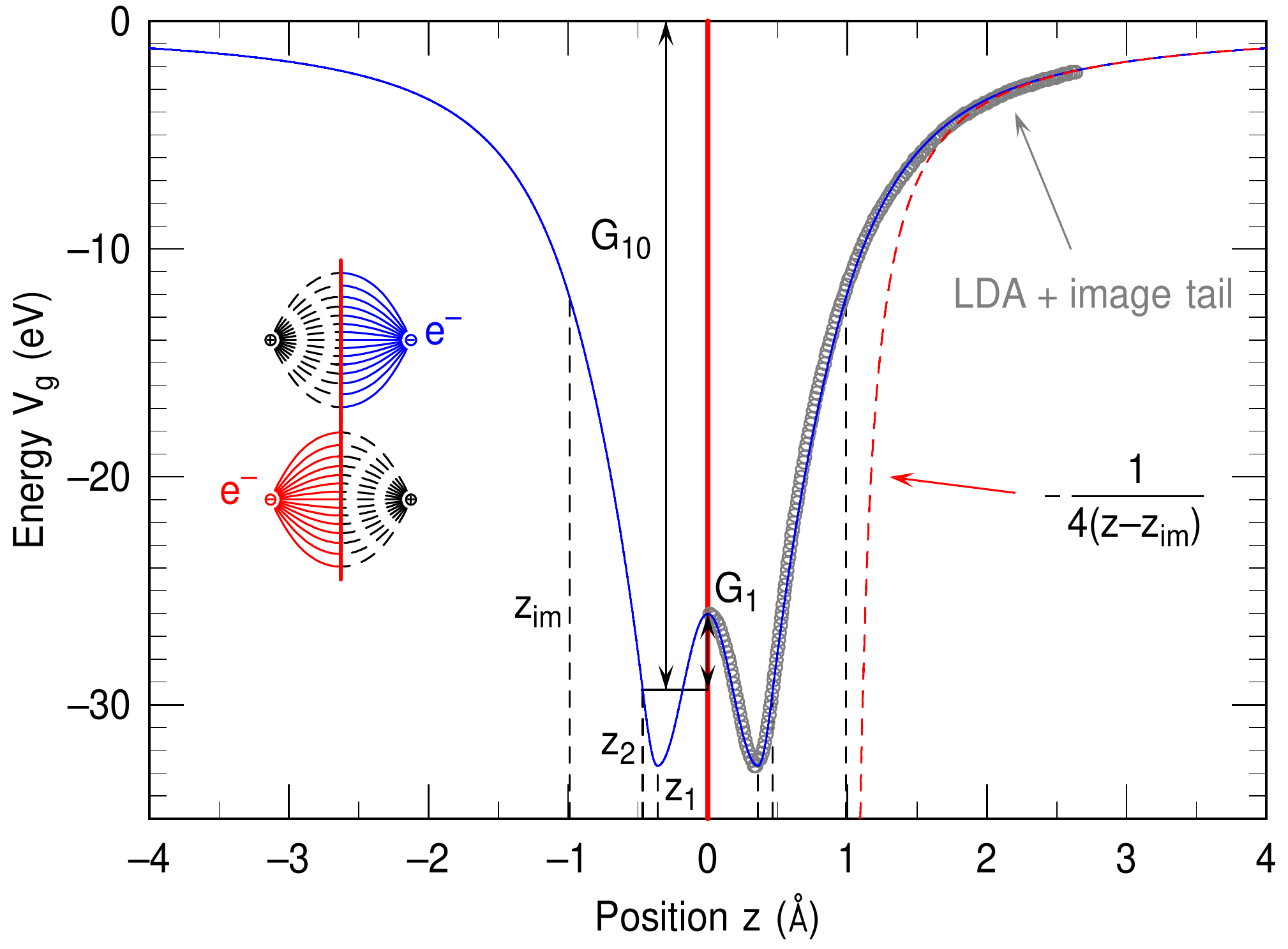}
        \caption[]{One-dimensional model-potential of freestanding
            graphene.
             The vertical dashed lines delimit the different
            definition intervals.
             The red dashed line represents the asymptotical form of
            the classical potential due to the image-charges of both
            surfaces (inset).
             Gray circles depict the "LDA+image tail" potential
            which has been extracted from
            figure~1 of \cite{Silkin09prb}.
            \label{fig_g_pot}}
    \end{indented}
\end{figure}
 Our analytical approximation of the "LDA+image tail" hybrid
potential of Silkin \emph{et al.} \cite{Silkin09prb} for the
description of freestanding graphene is inspired by the definition
of the metal potential with omission of the bulk part:
\begin{equation}
V_{\rm g}(z) = \cases{
    -G_{10} +G_{1}\cos(\beta_{1}|z|)                        &$0 < |z| \leq |z_{1}|$\\
    -G_{20} -G_{2}\cos[\beta_{2} (|z|-|z_{1}|)]  &$|z_{1}| <  |z| \leq |z_{2}|$\\
    -G_{3}\exp[-\alpha(|z|-|z_{2}|)]                                   &$|z_{2}| < |z| \leq |z_{\rm im}|$\\
    \;\;\,\frac{\exp[-\lambda(|z|-|z_{\rm im}|)]-1}{4(|z|-|z_{\rm im}|)} &$|z_{\rm im}|<  |z|$.
    }\label{eq_g_pot}
\end{equation}

 As shown in \fref{fig_g_pot}, $V_{\rm g}(z)$ is symmetric in the
$z$-direction with respect to the graphene layer at $z=0$ and
converges to the classical image-potential for large distances
$|z|$.
 Again, only four of the twelve parameters $z_{1}$, $z_{2}$, $z_{\rm
im}$, $G_{1}$, $G_{10}$, $G_{2}$, $G_{20}$, $G_{3}$, $\beta_{1}$,
$\beta_{2}$, $\alpha$, and $\lambda$ are independent if we require
$V_{\rm g}(z)$ to be continuously differentiable at the matching
points $z_{1}$, $z_{2}$, and $z_{\rm im}$.
 We choose the potential offset $G_{10}=\Phi_{g}+V_{0}$, the
amplitude $G_{1}$, and the inverse widths $\beta_{1}$ and
$\beta_{2}$ to fit $V_{\rm g}(z)$ to the "LDA+image tail" hybrid
potential for a matching point between LDA potential and image-tail
of $z_0=3$~a.u. The best fit is shown as solid line in
\fref{fig_g_pot}.
 It has been achieved with $G_{1}=3.336$~eV, $G_{10}=29.3465$,
$\beta_{1}=8.7266$ and $\beta_{2}=14.8353$.

 The binding energies $E^{\pm}_{n}$ of the symmetric and
antisymmetric image-potential states calculated with these
parameters are listed in \tref{tab_g_pot}.
 They agree well with those calculated using the "LDA+image tail"
hybrid potential for $z_0=3$~a.u. \cite{Silkin09prb}
 The two equivalent positions of the image-planes of $z_{\rm im}=\pm
0.99$~{\AA}, which result from the chosen parameters, are almost
coincident with the spatial extent of the polarizable conjugated
$\pi$-system \cite{Silkin09prb}.

%
\Table{Binding energies $E_{n^\pm}$~(eV) calculated with the
        one-dimensional model potential and the
        "LDA+image tail" hybrid potential \cite{Silkin09prb},
        respectively.\label{tab_g_pot}}
\br
 & $E_{1^{+}}$ & $E_{1^{-}}$ & $E_{2^{+}}$ & $E_{2^{-}}$\\
\mr
1D-Potential          &  1.46 & 0.60 & 0.28 &  0.18 \\
LDA+image tail & 1.47 &  0.72 & 0.25 & 0.19 \\
\br
\endTable

 The screening of electric fields by the metal as well as the
graphene layer is taken into account by cutting the corresponding
image tail of $V_{\rm m}(z)$ and $V_{\rm g}(z)$ at the respective
opposite surface ($V_{\rm m}(z)=0$ for $z>d_{\rm g}$ and $V_{\rm
g}(z)=0$ for $z<0$).
 Together with the corrections $\delta V(z)$ and $V_{\rm \Phi}(z)$,
this results in a discontinuity of the total model potential $V(z)$
at $z=0$ and $z=d_{\rm g}$.
 The discontinuity is compensated by increasing the amplitudes of the
respective cosine oscillation of the metal potential $V_{\rm m}$
\eref{eq_mpot} and the graphene potential $V_{\rm g}$
\eref{eq_g_pot}.

\section*{References}                                              %

\end{document}